\documentclass[12pt]{article}
\newcommand {\numucc} {$\nu_\mu$CC}
\newcommand {\anumucc} {$\bar{\nu}_\mu$CC}

\newcommand {\lazene} {$\Lambda K^0_S \pi^-$}
\newcommand {\lazepo} {$\Lambda K^0_S \pi^+$}
\newcommand {\lazecha} {$\Lambda K^0_S \pi^\pm$}
\newcommand {\lanene} {$\Lambda K^- \pi^-$}
\newcommand {\lanepo} {$\Lambda K^- \pi^+$}
\newcommand {\lanecha} {$\Lambda K^- \pi^\pm$}

\newcommand {\lamkapi} {$\Lambda K \pi$}
\newcommand {\trimass} {$m(\Lambda K \pi)$}
\newcommand {\thetano} {$\theta_\mathrm{norm}$}
\newcommand {\abscosno} {$|\cos\theta_\mathrm{norm}|$}
\newcommand {\acropla} {$|\cos\theta_\mathrm{norm}| > 0.5$}
\newcommand {\acro} {$|\cos\theta_\mathrm{norm}| > 0.7$}

\begin {document}
\title {Search for formation of isospin-3/2 $\Xi$ states by neutrinos} 

\author { A.E. Asratyan \\
{\normalsize\it Institute of Theoretical and Experimental Physics, Moscow, 117218 Russia}\\
{\normalsize E-mail: ashot.asratyan@gmail.com } }

\date {\today}
\maketitle

\begin{abstract}
     A narrow peak near 1870 MeV is observed in the 
combined invariant-mass spectrum of the systems 
$\Lambda K^0_S \pi^-$, 
$\Lambda K^0_S \pi^+$, 
$\Lambda K^- \pi^-$, and
$\Lambda K^- \pi^+$
formed in $\nu_\mu$- and $\bar{\nu_\mu}$-induced charged-current collisions 
with free protons, deuterons, and Neon nuclei. Observed width of the resonance 
is consistent with being entirely due to apparatus smearing. A possible 
interpretation of the peak is formation and three-body $\Lambda K \pi$ decay 
of an exotic baryon with $I = 3/2$ and $S = -2$. 
\end{abstract}

     A narrow peak near 1862 MeV has been observed in invariant masses 
of the $\Xi^-\pi^-$ and $\Xi^-\pi^+$ systems formed in $pp$ collisions 
\cite{na49}, tentatively interpreted as a baryon state with $S = -2$
and $I = 3/2$ that is part of the (hypothesized) antidecuplet of
pentaquark baryons \cite{theory}. Here, I report on a search for
formation of this exotic baryon in neutrino and antineutrino collisions
with free protons, deuterons, and neon nuclei. Instead of $\Xi^-\pi$,
I analyze the system \lamkapi\ which may provide access to all four 
charge states of the $\Xi_{\bar{10}}$: 
$\Xi_{\bar{10}}^+ \rightarrow \Lambda \bar{K^0} \pi^+$,
$\Xi_{\bar{10}}^0 \rightarrow \Lambda K^- \pi^+$,
$\Xi_{\bar{10}}^- \rightarrow \Lambda \bar{K^0} \pi^-$, and
$\Xi_{\bar{10}}^{--} \rightarrow \Lambda K^- \pi^-$.
The system \lamkapi\ has a higher mass threshold than $\Xi\pi$, but on 
the other hand the smallness of observed $\Xi_{\bar{10}}$ width indicates 
that the kinematically favored decay $\Xi_{\bar{10}} \rightarrow \Xi \pi$
is subject to some dynamic suppression which may render the three-body 
\lamkapi\ channel competitive.

     As in the previous search for formation of the $\Theta^+(1540)$ baryon 
\cite{myself}, I analyze the data collected by several neutrino experiments 
with big bubble chambers---BEBC at CERN and the 15-foot chamber at Fermilab. 
Though logged several decades ago, bubble-chamber neutrino data are still 
unrivaled in quality and completeness. I rely on a database that comprises 
some 120 000 $\nu_\mu$- and $\bar{\nu}_\mu$-induced charged-current (CC) events 
on hydrogen, deuterium, and neon targets. In the past, these combined 
bubble-chamber neutrino data were employed in a number of physics analyses 
\cite{combine}. The database embraces the bulk of neutrino data obtained with 
BEBC (experiments WA21, WA25, and WA59) and a significant fraction of those 
collected with the 15-foot bubble chamber (experiments E180 and E632). 
\begin{table}[!t]

\small
\begin{tabular}{|l|c|c|c|c|c|c|}
\hline
Experiment          &  WA21 &  WA25 &  WA59 &  E180 & E632 &Total\\
Chamber                     &BEBC   &BEBC   &BEBC &15' B.C.&15' B.C.& \\
Fill      &Hydrogen &Deuterium &Neon--H$_2$ &Neon--H$_2$ &Neon--H$_2$& \\
\hline
\hline
{\bf Neutrinos:}      &       &       &       &       &      &  \\
Mean $E_\nu$, GeV     &48.8   &51.8   &56.8   &52.2   &136.8 &56.7\\
Mean $K^0_S$($\Lambda$)&5.7(3.5)&5.7(3.4)&4.5(2.8)&3.4(1.9)&7.7(5.1)&5.3(3.3)\\
\ \ \ momentum, GeV&      &       &      &       &      &\\
All CC events         &18746  &26323  & 9753  & 882 &8550(5621) &64250\\
CC events with $K^0_S$ &1050  & 1279  &  561  &  21 & 587      &3498 \\
CC events with $\Lambda$ &442 &644  & 378   & 19   & 352     &1835 \\
CC events with           & 41 & 76  & 46    & 0    & 52      & 215 \\
\ \ \ $\Lambda$ and $K^0_S$ &  &    &       &      &         &     \\ 
\hline
{\bf Antineutrinos:}   &       &       &       &       &   &    \\
Mean $E_\nu$, GeV      &37.5  &37.9    &39.5   &33.8 &110.0 &38.9 \\
Mean $K^0_S$($\Lambda$)&4.2(2.5)&4.2(2.0)&3.5(2.1)&3.4(1.4)&7.6(2.9)&4.0(2.1)\\
\ \ \ momentum, GeV &  &   &       &     &      &  \\
All CC events         &13155  &16314  &15693  &5927  &1810(1190) &52900 \\ 
CC events with $K^0_S$ &702   &761    &631    &231   &123       &2448  \\
CC events with $\Lambda$ &427 &459   &587    &165    &62      &1700 \\
CC events with           & 56 & 62  & 58    & 17    & 6      & 199 \\
\ \ \ $\Lambda$ and $K^0_S$ &  &    &       &      &         &     \\ 
\hline
\end{tabular}
\caption
{Relevant characteristics of the bubble-chamber neutrino data analyzed
in this paper. For E632, I show either the actual number of measured CC 
events (in the parentheses) and the ``equivalent" number that includes all 
CC events analyzed for $V^0$ emission.}
\label{statistics}
\end{table}
Total numbers and mean energies of \numucc\ and \anumucc\ events detected and 
reconstructed by the aforementioned experiments \cite{experiments} are 
summarized in Table \ref{statistics}. Also shown are the statistics of CC 
events with reconstructed $K^0_S$ mesons and $\Lambda$ hyperons in the final 
state. 

     The bubble chamber is a good spectrometer, but provides virtually no
identification for charged kaons. (Still, a few are identified by bubble 
density, range consistent with track curvature, and decay signature at endpoint.)
\begin{figure}[!b]

\vspace{9 cm}
\includegraphics{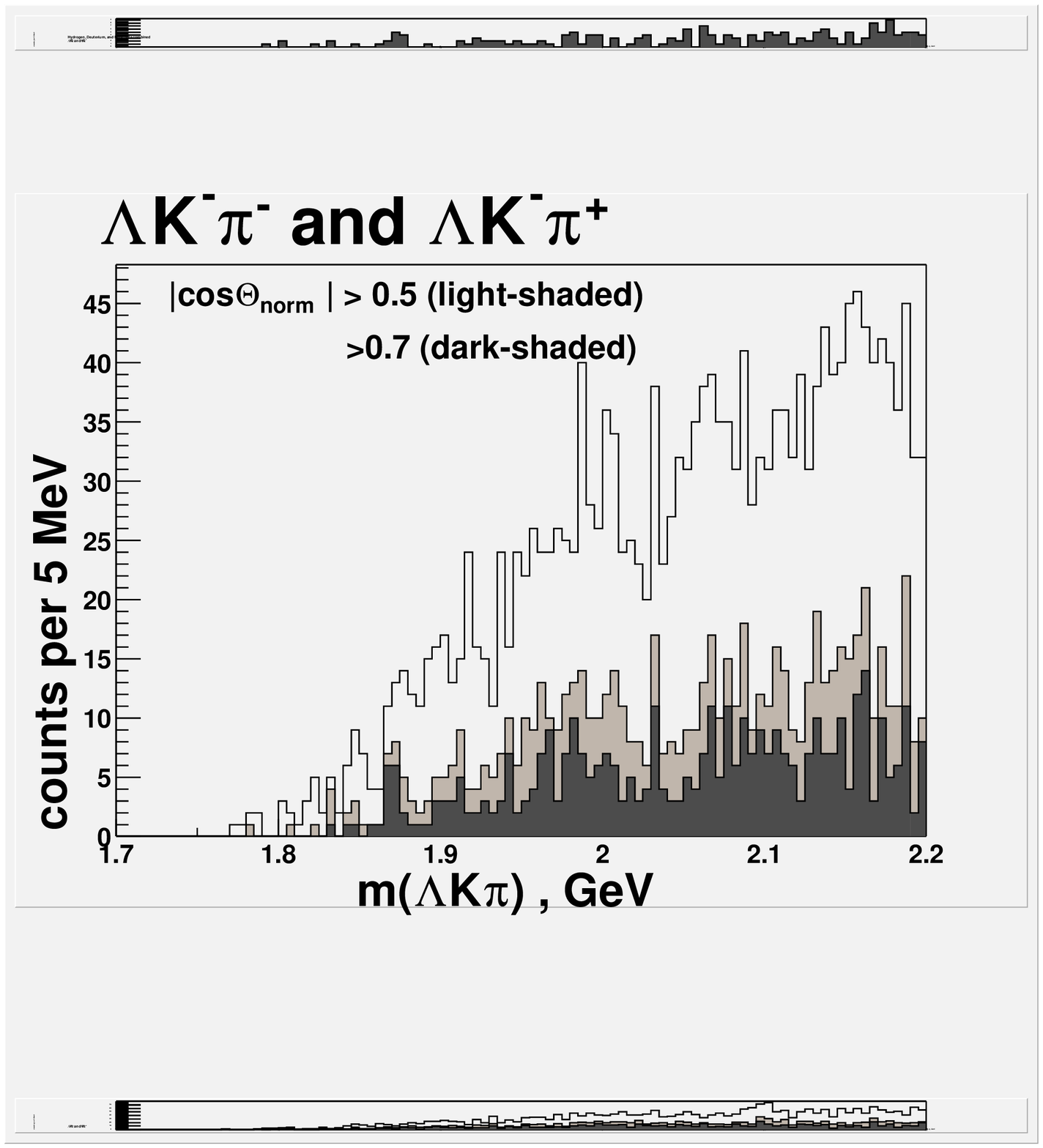}
\caption
{Invariant mass of the \lanene\ and \lanepo\ systems combined. The light- and 
dark-shaded histograms result from the selections \acropla\ and \acro, 
respectively.}
\label{abscosno}
\end{figure}
Therefore, kaon mass is combinatorially assigned to any negative hadron for which
the $K^-$ hypothesis was not ruled out at the stage of kinematic reconstruction.
I reject those $\Lambda K^-$ subsystems that fall in the $\Sigma^-(1385)$ mass 
region as soon as the pion hypothesis is selected: $1355<m(\Lambda\pi^-)<1415$ MeV.
The masses of all selected \lanecha\ systems are plotted in Fig.~\ref{abscosno}.
(Here and in what follows, I combine the neutrino and antineutrino data and 
those for all targets.) Despite the proximity of the \lamkapi\ mass threshold, 
including ``assigned" $K^-$ mesons is seen to result in a high level of 
combinatorial background. So I cut on an angle appropriate for 3-body decays, 
\thetano. In the \lamkapi\ frame, the 3-momenta of the three daughters lie in the 
same decay plane, and \thetano\ is defined as the angle between the normal to 
this plane and the \lamkapi\ boost direction from lab. (Note that 
$\cos \theta_\mathrm{norm} = \pm 1$ corresponds to exactly transverse position 
of the decay plane with respect to the \lamkapi\ direction of motion.) Given an 
unpolarized parent, the signal should be uniformly distributed in \abscosno. On 
the other hand, the mean value of \abscosno\ does not exceed 0.29 for all 
selected \lanecha\ systems, since inclusive hadrons are largely emitted with 
small transverse momenta to the hadron jet. The effects of the selections \acropla\ 
and \acro\ on the \lanecha\ mass spectrum are shown in Fig.~\ref{abscosno}. Note
that in a narrow region near 1870 MeV, the mass spectrum is less depleted by 
cutting on \abscosno\ than in the upstream and downstream regions. Since $K^0_S$ 
mesons are reliably identified by $K^0_S \rightarrow \pi^+ \pi^-$ decays, no
\abscosno\ selection is applied to the \lazene\ and \lazepo\ systems.

\begin{figure}[!b]

\vspace{12 cm}
\includegraphics{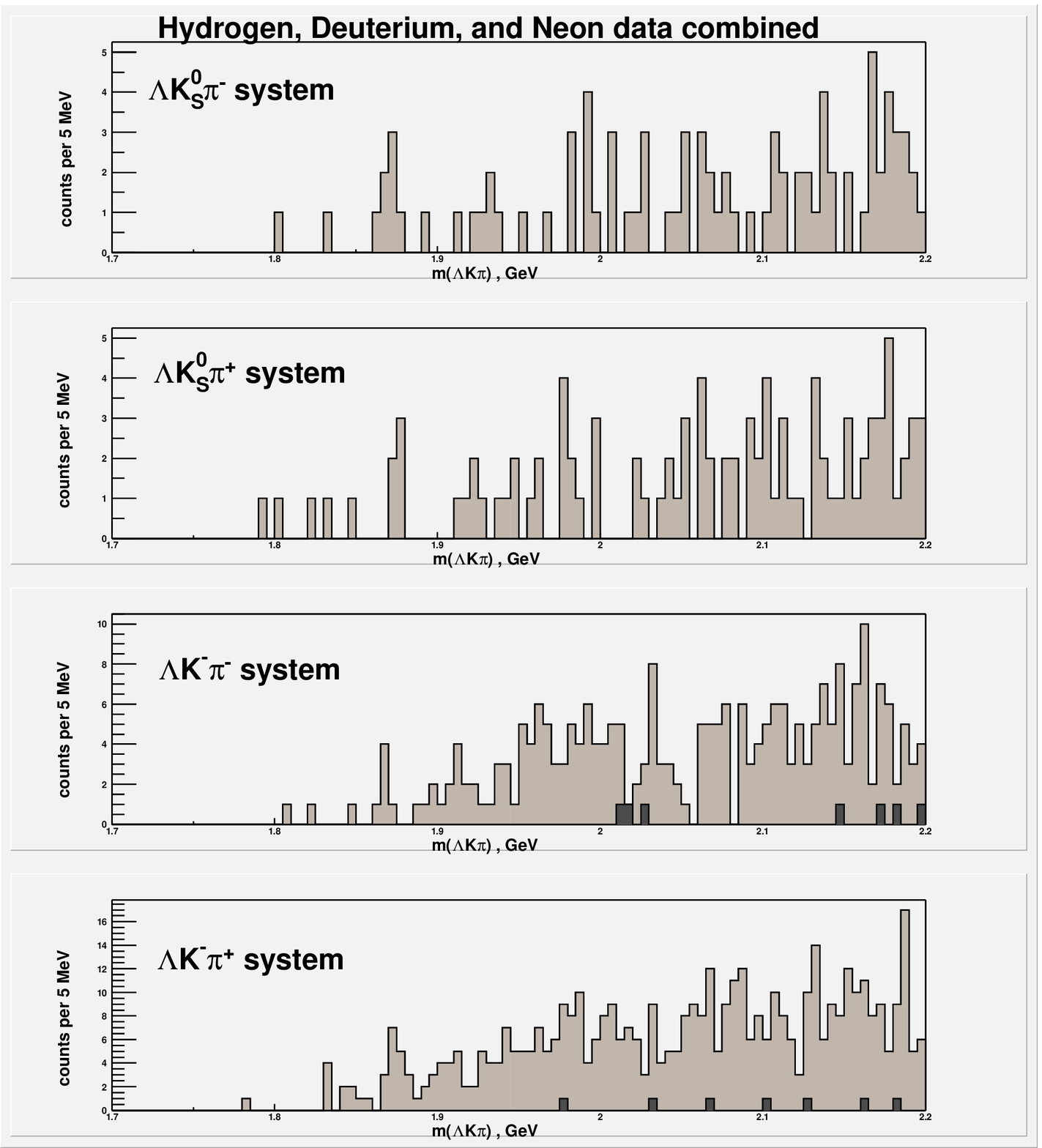}
\caption
{The \lazene, \lazepo, \lanene (\acropla)  and \lanepo\ (\acropla) mass 
distributions. Dark-shaded histograms are for identified charged kaons.}
\label{break-by-charge}
\end{figure}
\begin{figure}[!b]

\vspace{12 cm}
\includegraphics{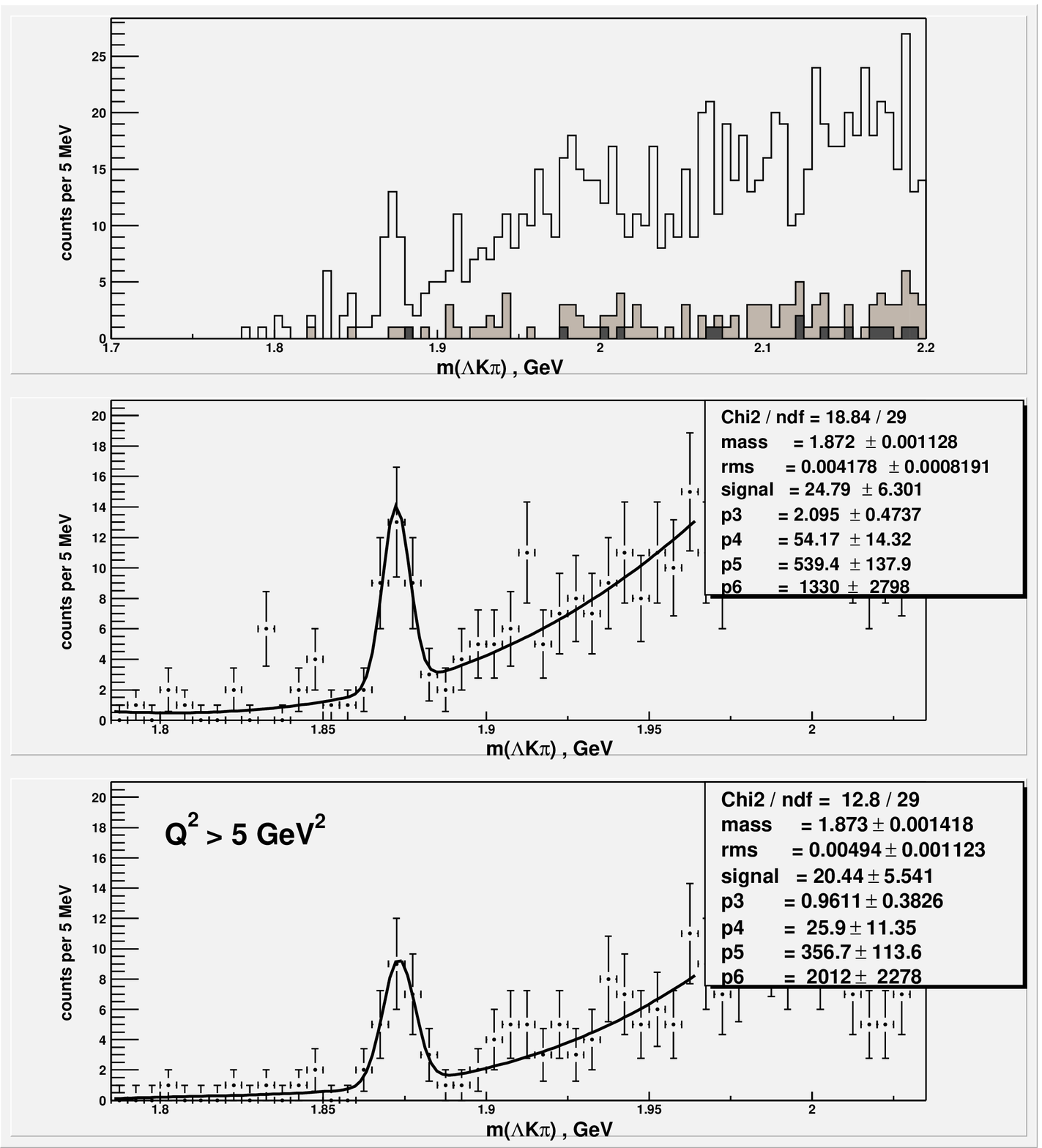}
\caption
{The \lazene, \lazepo, \lanene (\acropla), and \lanepo\ (\acropla) mass spectra 
added up for all Ne+H$_2$+D$_2$ data (top panel). The light- and dark-shaded
areas are the contributions from events with additional $K^0_S$ mesons and 
identified charged kaons, respectively. Shown in the middle panel is a Gaussian 
fit of the combined \trimass\ spectrum. The effect of an additional selection 
$Q^2 > 5$ GeV$^2$ is illustrated in the bottom panel.}
\label{combined}
\end{figure}
\begin{figure}[!b]

\vspace{12 cm}
\includegraphics{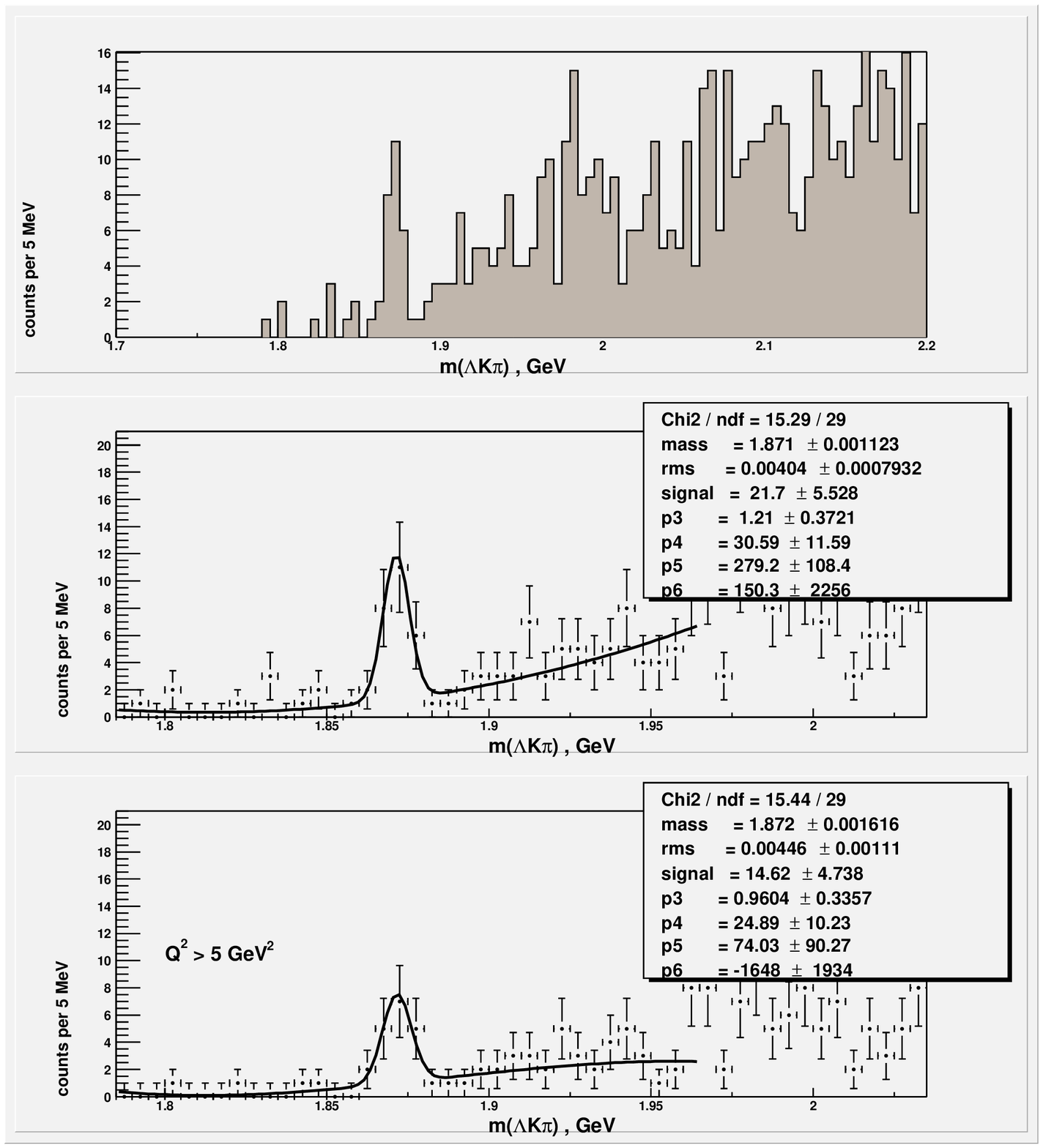}
\caption
{Similar data as in Fig.~\ref{combined}, but with a tighter selection \acro\ for
the \lanene\ and \lanepo\ systems.}
\label{combi}
\end{figure}

     Invariant masses of selected \lazene, \lazepo, \lanene\ (\acropla), and 
\lanepo\ (\acropla) systems are separately plotted in Fig.~\ref{break-by-charge}. 
All four show small enhancements near 1870 MeV. And finally, in Figs. 
\ref{combined} (\acropla) and \ref{combi} (\acro) I add up the mass spectra for 
all selected \lazecha\ and \lanecha\ systems, neglecting possible mass differences 
between the states of different charges. A distinct narrow enhancement is seen at 
$m(\Lambda K \pi) \simeq 1870$ MeV. The ``grand-total" \trimass\ spectrum is then 
fitted to a Gaussian on top of a third-order polynomial, see the middle panels of 
Figs. \ref{combined} and \ref{combi}. Either fit returns a central mass value
slightly in excess of 1870 MeV and an rms width of $\sigma \simeq 4\pm1$ MeV. The 
observed width is consistent with being entirely due to apparatus smearing of
\trimass, estimated as $\sim 5$ MeV using individual errors for live events in 
the peak region. Statistical significance of the putative signal, (optimistically)
estimated as $S/\sqrt{B}$ over the mass region of $\pm 2\sigma$ around the peak
position, is over 8 standard deviations.

     Two events in the peak have $K^0_S$ mesons among the secondaries emitted in 
association with the \lamkapi\ system, and yet another one --- an associated  
charged kaon which is a $K^+$ identified in neon, see Fig.~\ref{combined}. Had 
two $s\bar{s}$ pairs been produced per (anti)neutrino collision, one would 
expect $\sim 6\pm3$ events with associated $K^0_S$ mesons from fragmentation
of the two $\bar{s}$ quarks. Note however that two $s$ quarks may also result 
from a strangeness-changing transition $u \rightarrow s$ accompanied by 
creation of a single $s\bar{s}$ pair. 

     In summary, a narrow peak near 1870 MeV is observed in the combined 
invariant-mass spectrum of the systems \lazene, \lazepo, \lanene, and \lanepo\ 
formed in $\nu_\mu$- and $\bar{\nu_\mu}$-induced CC collisions with free 
protons, deuterons, and Neon nuclei. Observed width of the putative \lamkapi\ 
resonance is consistent with being entirely due to apparatus resolution. A 
possible interpretation of the peak is formation and \lamkapi\ decay of an 
exotic baryon with $I = 3/2$ and $S = -2$. Our results may support the 
earlier observation of a $\Xi^- \pi^\pm$ resonance near 1862 MeV in $pp$ 
collisions \cite{na49}, provided that the discrepancy of $\sim$10 MeV 
between the masses of the two resonances can be explained by systematic 
effects.

     I thank the WA21, WA25, WA59, E180, and E632 Collaborations for 
providing excellent neutrino data whose physics potential is still far from 
exhausted. Thanks are also due to Prof. Ya.I. Azimov and Dr. I.I. Strakovsky 
for useful comments and stimulating discussions.

\newpage

\end{document}